\documentclass[aps,prb,10pt,twocolumn,showpacs,superscriptaddress,floatfix]{revtex4-2}

\usepackage{graphicx}
\usepackage{dcolumn}
\usepackage{bm}
\usepackage{amsmath}
\usepackage{amssymb}
\usepackage{tabularx}
\usepackage{graphicx}
\usepackage{dcolumn}
\usepackage{bm}
\usepackage{amssymb}
\usepackage{placeins}
\usepackage{siunitx}
\usepackage{textalpha}
\usepackage{xcolor}
\usepackage{soul}
\usepackage[bookmarks=false]{hyperref}
\usepackage[export]{adjustbox}
\usepackage{hyperref}
\usepackage{cleveref}

\newcommand{\blue}[1]{{\color{blue}#1}}
\FloatBarrier

\begin{document}

\title{Anisotropic metamagnetism and magnetotransport of heavy rare-earth orthorhombic single-crystal TbAlGe}

\author{Ram Kumar}
\email{ramphy21@umd.edu}
\affiliation {Maryland Quantum Materials Center, Department of Physics, University of Maryland, College Park, MD 20742, USA}

\author{K.E. Avers}
\affiliation {Maryland Quantum Materials Center, Department of Physics, University of Maryland, College Park, MD 20742, USA}
\author{V. Saini}
\affiliation {National High Magnetic Field Laboratory, MS E536, Los Alamos National Laboratory, Los Alamos, NM, 87545, USA}
\author{D. S. Sokratov}
\affiliation {Maryland Quantum Materials Center, Department of Physics, University of Maryland, College Park, MD 20742, USA}
\author{Y. Anand}
\affiliation {Maryland Quantum Materials Center, Department of Physics, University of Maryland, College Park, MD 20742, USA}
\author{P. Saraf}
\affiliation {Maryland Quantum Materials Center, Department of Physics, University of Maryland, College Park, MD 20742, USA}
\author{J. A. Horn}
\affiliation {Maryland Quantum Materials Center, Department of Physics, University of Maryland, College Park, MD 20742, USA}
\author{N. Brenowitz}
\affiliation {Maryland Quantum Materials Center, Department of Physics, University of Maryland, College Park, MD 20742, USA}
\author{S. Otazo}
\affiliation {Maryland Quantum Materials Center, Department of Physics, University of Maryland, College Park, MD 20742, USA}
\author{P. Sobel}
\affiliation {Maryland Quantum Materials Center, Department of Physics, University of Maryland, College Park, MD 20742, USA}
\author{D. Graf}
\affiliation {National High Magnetic Field Laboratory, Tallahassee, Florida 32310, USA}
\author{S. R. Saha}
\affiliation {Maryland Quantum Materials Center, Department of Physics, University of Maryland, College Park, MD 20742, USA}
\author{J. Paglione}
\email{paglione@umd.edu}
\affiliation {Maryland Quantum Materials Center, Department of Physics, University of Maryland, College Park, MD 20742, USA}
\affiliation {Canadian Institute for Advanced Research, Toronto, ON M5G-1Z8, Canada}

\date{\today}

\begin{abstract}
We report a comprehensive investigation of the anisotropic magnetism and magnetic field-induced transitions in single crystals of the orthorhombic system TbAlGe, a member of the topological RAlGe (R = rare-earth) family with the highest ordering temeprature in the RAlX (X = Si, Ge) series. 
With a single rare earth site with triangular coordination in its $Cmcm$ orthorhombic unit cell, TbAlGe harbors complex magnetic interactions that yield two antiferromagnetic transitions at 40~K and 8~K in zero field, and a rich cascade of metamagnetic transitions that only appear for fields directed along the crystallographic $a$-axis. 
Combining electrical resistivity, magnetization and heat capacity measurements with magnetotransport experiments performed up to 41.5~T, we construct a magnetic phase diagram mapping the multiple magnetic phases of TbAlGe, and discuss  the complex interplay between localized 4$f$ magnetism and itinerant electronic topology, establishing TbAlGe as a compelling platform for exploring tunable magnetic semimetal physics.

\end{abstract}

\maketitle

\section{Introduction}
Weyl semimetals exhibit a variety of remarkable physical properties, including topologically protected electronic states that coexist with conventional metallic conduction, giving rise to unconventional transport phenomena. 
Recently, magnetic topological materials in the family of strongly correlated \textit{f}-electrons intermetallics have attracted much attention as they provide rich physical properties due to the interplay between correlated-electron phenomena and topological electronic bands \cite{Bernevig2022}, which make them valuable for both fundamental research and technological applications \cite{KHJBuschow_1977,RevModPhys.81.1551}. More recently, the rare-earth based RAlGe (R = rare earth) intermetallics have attracted considerable attention as potential Weyl semimetal candidates. This family can host either type I or type II Weyl fermions by breaking time-reversal symmetry or inversion symmetry, or both simultaneously, depending on suitable choice of the rare-earth species \blue{\cite{PhysRevB.105.174502, PhysRevB.103.165128, doi:10.1126/sciadv.1603266, PhysRevB.98.245132, PhysRevLett.124.017202, PhysRevB.103.214401}}. 
Interestingly, this series can crystallize in both the tetragonal $\alpha$-ThSi$_2$ type structure for light rare earths, and the orthorhombic YAlGe-type structure for heavy rare earths with space group $Cmcm$ \cite{PhysRevLett.124.017202, WANG2021167739, WANG2022163623}. Although the structures appear different at first glance, they both have the same motif of the 'R' being enclosed by a triangular prism. The structure with $\alpha$-ThSi$_2$ type has a triangular prism with [R$_6$] as the vertices, and the YAlGe type is a triangular prism with [R$_2$Al$_4$] as the vertices. With a similar triangular prism, the main feature of the YAlGe type structure is that the R ions form a buckled triangular lattice in the a-c plane, which may lead to anisotropic exchange interactions in the system.

\begin{figure*}[htbp]
	\centering
    \adjincludegraphics[height=9cm,trim={0cm 1cm 0cm 0cm},clip]{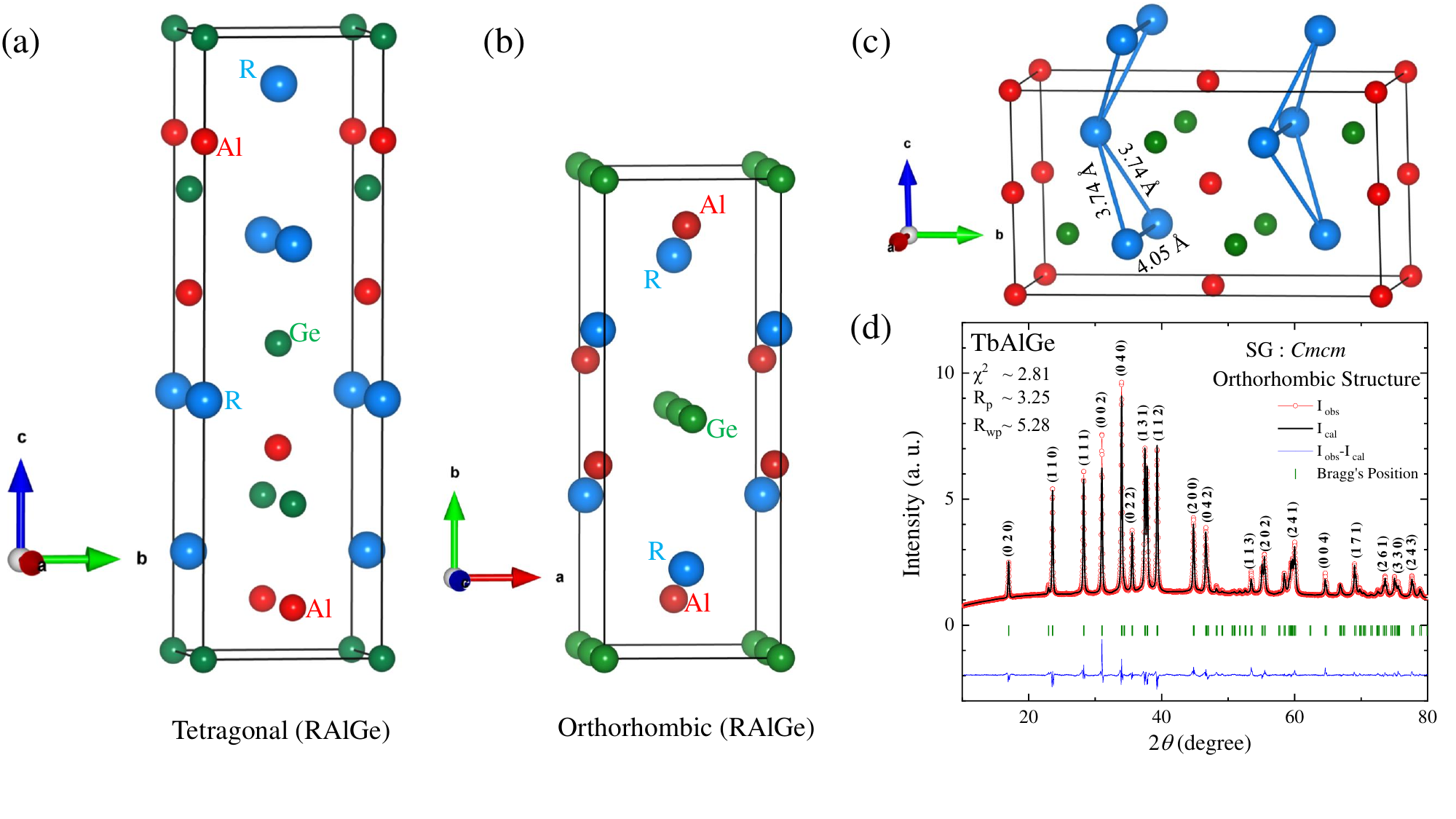}
	\caption{A comparison of two types of crystal structure of RAlGe, tetragonal (a), and orthorhombic (b). Panel (c) shows the crystal structure with the distorted triangular arrangements of Tb-atoms in the orthorhombic crystal structure. Panel (d) represents Rietveld refinement of powder X-ray diffraction data with the experimental data (Red symbols) and the black line representing the calculated data. A set of vertical green bars represents the Bragg peak positions of the orthorhombic (Space group: $Cmcm$) structure, and the blue line at the bottom represents the difference between the observed and calculated diffraction patterns.}
	\label{fgr:XRD}
\end{figure*}


A majority of the RAlGe family exhibits unique magnetic and transport characteristics, often linked to their novel electronic and magnetic structures. RAlGe (R= Ce and Pr) show interesting magnetic and electric properties \cite{PhysRevB.98.245132, PUKAS2004162, FLANDORFER1998191,doi:10.1126/science.aat0348,Singh02072020, 10.1063/1.5090795, Destraz2020, Sanchez2020, 10.1063/1.5132958}. The physical properties of single crystalline NdAlGe shows field-induced metamagnetic transition \cite{Zhao_2022}, large anomalous Hall effect, and helical magnetic structure \cite{PhysRevB.107.224414, PhysRevMaterials.7.034202}. GdAlGe is found to have two structures, i.e., a high temperature (HT) phase with tetragonal LaPtSi type (I4$_1$md) and a low temperature (LT) phase with orthorhombic YAlGe type ($Cmcm$) \cite{Wang_2020, WANG2021167739, WANG2022163623}. Less attention has been given to the heavy rare-earth systems; to the best of our knowledge, the other heavy rare-earth RAlGe intermetallics have only been investigated in their polycrystalline form \cite{doi:10.1021/jpcc.1c07073, WANG2021167739, WANG2022163623}. 
GdAlGe and TbAlGe are only two single crystal studies of the heavy rare-earth side. Orthorhombic GdAlGe, which is at the transition point between tetragonal and orthorhombic phases in the rare earth series, has been shown to have Dirac-like linear band dispersion \cite{r7gz-kwqw}. On the other hand in TbAlGe, the preliminary study on polycrystalline  reports an antiferromagnetic (AFM) transition below 40 K \cite{doi:10.1021/jpcc.1c07073}; while magnetization has only been measured in a single crystal only upto 7 T \cite{Gouda_2022}. 

In this study, we present the first comprehensive investigation of magneto-transport properties of single crystalline orthorhombic TbAlGe up to magnetic fields as high as 41.5~T. Using magnetotransport, magnetization and heat capacity measurements performed on single crystals, we characterize the evolution of two AFM states that transition at $T_{N1}$ = 40~K and $T_{N2}$ = 8~K at zero field, as they evolve with temperature and magnetic field to reveal a rich texture of multiple metamagnetic transitions only for fields along the crystallographic $a$-axis. We characterize this anisotropic response and discuss how the anisotropic magnetic exchange interactions play a central role in stabilizing the complex spin structures and the tunable landscape for interesting physics in this system.

\section{EXPERIMENTAL DETAILS and crystal Chemistry}
Single crystals of TbAlGe were grown using the high-temperature self-flux technique with molten Al as a solvent. Tb with 99.9 $\%$ purity, Al (shot) and Ge (lump) with purity greater than 99.999$\%$, were obtained from Alfa-Aesar. The reactions were carried out in 2-cm$^3$ alumina crucibles, which were encapsulated in evacuated fused silica jackets by flame sealing. The temperature profile: ramping to 1100$^{\circ}$C with the rate 100$^{\circ}$C/hr, homogenization for up to 24~hr, cooling to 850$^{\circ}$C at 2$^{\circ}$C/hr rate. At 850$^{\circ}$C, the molten Al was removed by centrifugation. The as-grown crystal is shown in the inset of Fig. ~\ref{fgr:XRD}(b). The phase purity of these crystals was checked with an X-ray diffractometer (Rigaku, MiniFlex600), where X-ray diffraction (XRD) patterns were recorded with monochromated Cu-\textit{K$\alpha$} radiation ($\lambda$$\sim$1.5406~$\AA$), and crystal orientation along principal crystallographic directions was determined by Laue diffraction using a polychromatic X-ray source.
The Rietveld refinement technique \cite{Rietveld:a07067} using FullProf software packages has been used for phase identification. Further, the chemical composition is verified using a scanning electron microscope (SEM) with an energy-dispersive X-ray spectroscopy (EDS). The magnetic measurements were carried out using a SQUID-VSM, Quantum Design, and Dynacool with VSM-mode for magnetic field (\textit{B} = $\mu$$_0$H) up to 14~T. For every magnetic measurement, the magnetic field was oscillated to zero before each measurement to minimize residual magnetic flux trapped in the superconducting solenoid. Thermodynamic and electrical transport measurements were performed in the range 1.8–300~K and up to 14~T using a Dynacool cryostat (Quantum Design). High-field magnetotransport measurements were performed using a single-axis rotation probe in a 41.5~T resistive magnet (Cell-6) at the National High Magnetic Field Laboratory, with a base temperature of $\sim$1.5~K provided by a variable temperature insert.

\begin{figure}[!htbp]
	\adjincludegraphics[height=15cm,trim={0.7cm 0.25cm 0.25cm 0.7cm},clip]{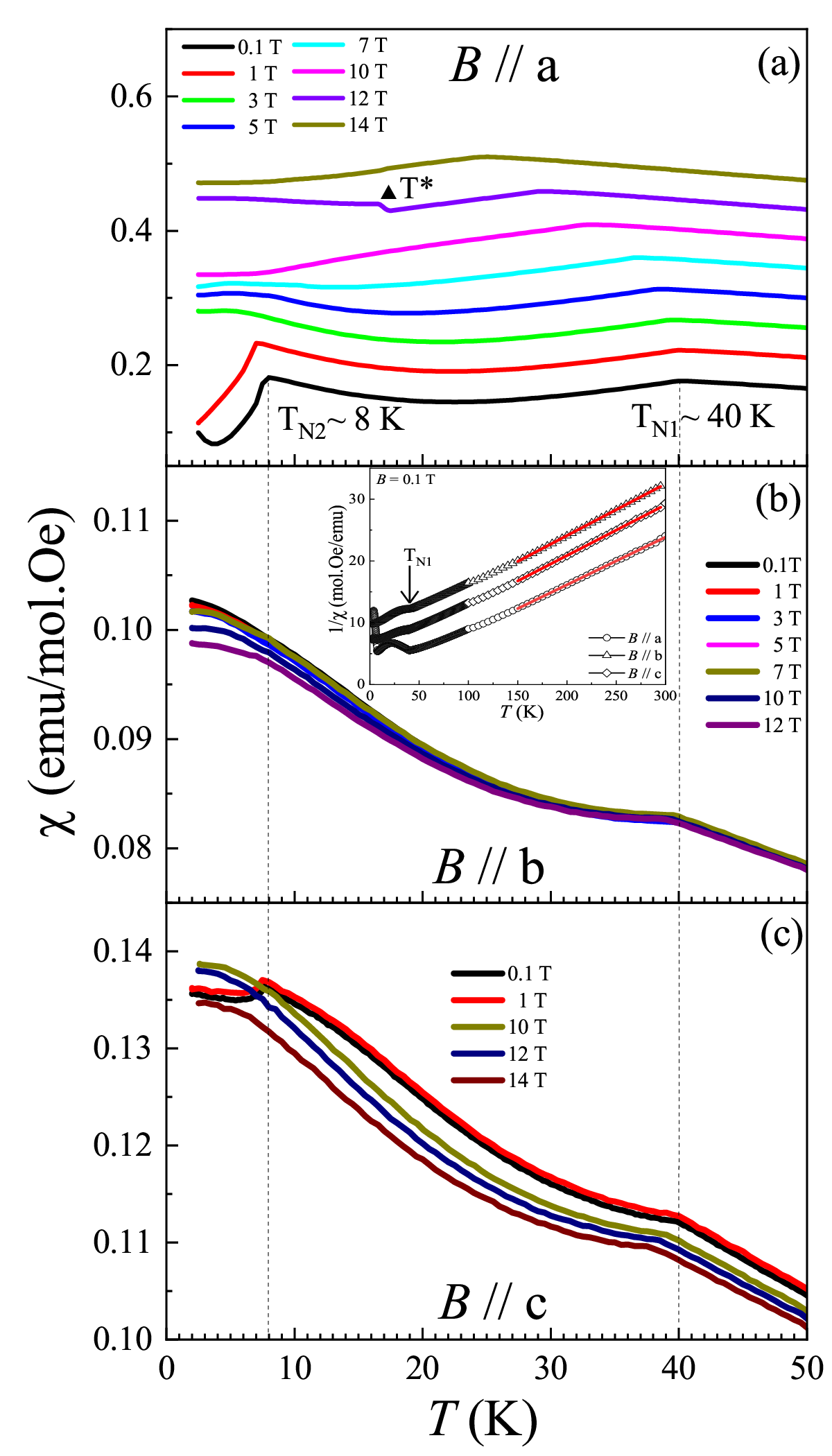}
\caption{Temperature dependence of magnetic susceptibility $\chi{_{dc}}$(\textit{T}) of TbAlGe measured under a magnetic field from 0.1~T to 14~T in all three $B \parallel a$, $B \parallel b$, and $B \parallel c$ directions as shown in fig.~2~(a), (b), and (c), respectively. The vertical dotted lines across all three panels mark the existing magnetic transitions at \textit{B} = 0.1~T. The curves in fig.~2~(a) are shifted upward for clarity to illustrate the magnetic transition under applied fields. A black triangle symbol is showing a field-induced transition (\textit{T$^*$}). Inset of (b) shows the Curie-Weiss fit. All data are measured on warming after cooling in a zero field.}
	\label{fgr:MvsT}
\end{figure}

\begin{figure}[!htbp]
	\adjincludegraphics[height=15cm,trim={0.7cm 0cm 4cm 0.35cm},clip]{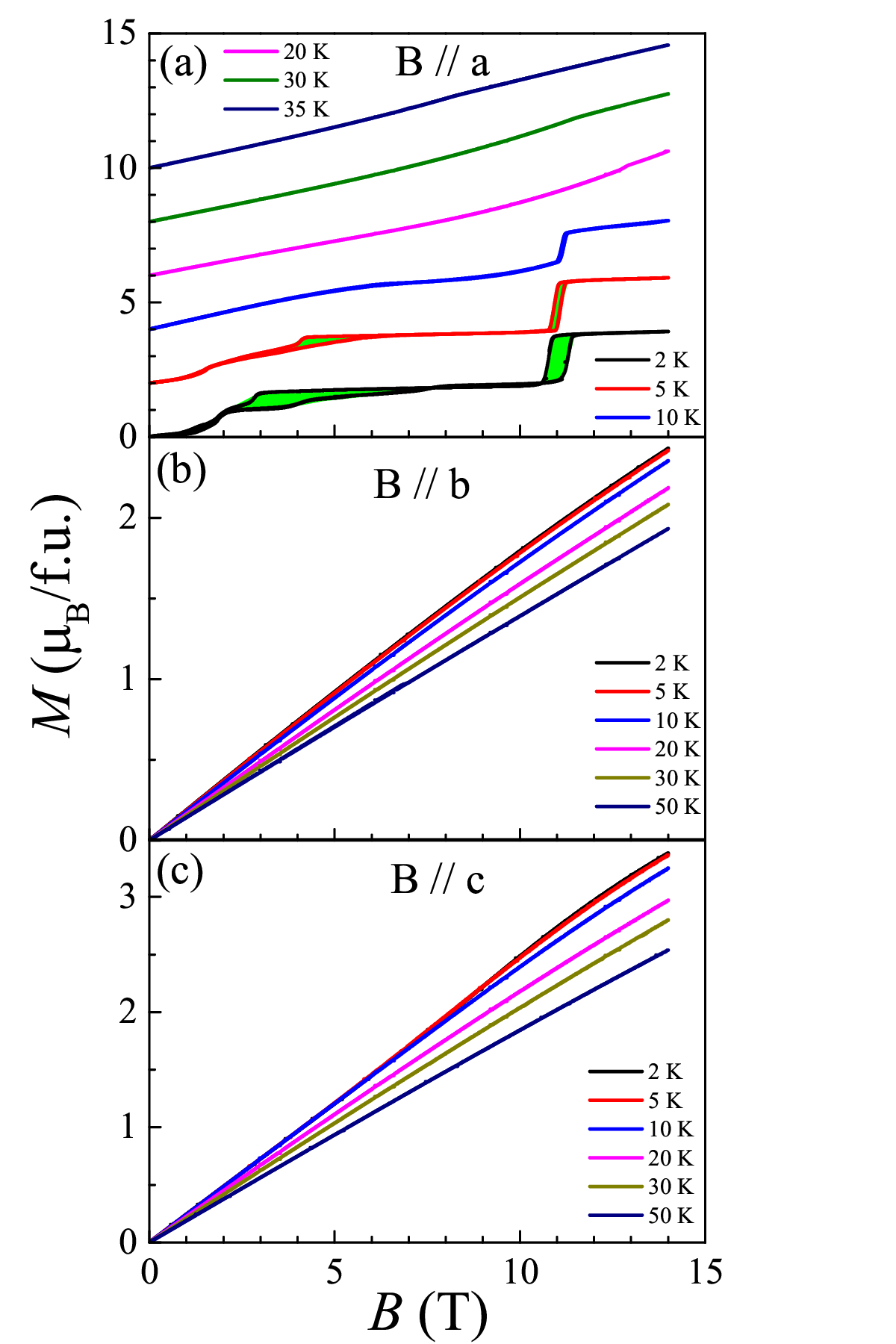}
    \caption{Magnetic field (\textit{B}) dependence of magnetization (\textit{M}) of TbAlGe measured at different temperatures for the applied magnetic field parallel to three principal crystallographic directions, i.e., (a) \textit{$B \parallel a$}, (b) \textit{$B \parallel b$}, and (c) \textit{$B \parallel c$}, respectively. We note that for \textit{$B \parallel a$}, there are multiple metamagnetic transitions, while in contrast, the other field directions are mostly featureless with minor curvature. The curves in fig.~3~(a) are shifted upward for clarity to illustrate the hysteresis at each metamagnetic transition.}
	\label{fgr:MvsB}
\end{figure}


\section{Results and Discussion}
\subsection{Structural properties}

Fig.~\ref{fgr:XRD}(a) shows the crystal structure of TbAlGe, which clearly depicts the triangular lattice of Tb-atoms with opposite slanting to the nearest triangular lattice, and Fig.~\ref{fgr:XRD}(b) the Rietveld refinement of the powder XRD data obtained after grinding the single crystals at room temperature. Structural analysis confirms our TbAlGe crystals are in the orthorhombic structural phase, and not the tetragonal phase shown by the lighter rare earth compounds of the RAlGe series \cite{doi:10.1021/ic00025a021}. Detailed powder XRD refinements converge in the space group $Cmcm$ with $a$- $b$- and $c$-axis lattice parameters 4.045(1)~Å, 10.439(1)~Å and 5.767(1)~Å, respectively, which correspond well to previous polycrystalline studies \cite{doi:10.1021/jpcc.1c07073}. EDS performed on the cleaned surfaces of the single crystal yielded an atomic composition of 31.66(2)\% Tb, 32.02(1)\% Al, and 36.32(1)\% Ge for TbAlGe. These values are in good agreement with the compositions obtained from XRD refinement.

\subsection{Magnetic properties}
Fig.~\ref{fgr:MvsT} presents the temperature-dependent \textit{dc}-magnetic susceptibility $\chi{_{dc}}(T)$ of TbAlGe measured in the zero-field cooled mode with different applied fields along the three principal crystallographic axes. It is observed that $\chi{_{dc}}$(\textit{T}) exhibits two AFM phase transitions at the Neel temperatures of \textit{$T_{N1}$}${\sim}$ 40 K and \textit{$T_{N2}$}${\sim}$ 8 K. The Curie–Weiss temperatures and effective magnetic moment of the Tb ions is estimated by fitting the inverse susceptibility (1/ $\chi$) with the modified Curie-Weiss law, $\chi$(T) =  $\chi$$_0$ + C/(T - $\theta$$_P$), in the paramagnetic region 150–300 K, as shown in the inset of Fig. 2(b). Here, $\chi$$_0$, C, and $\theta$$_P$ are the temperature-independent susceptibility, Curie constant, and paramagnetic Curie–Weiss temperature, respectively. The Curie-Weiss fits of inverse susceptibilities in the paramagnetic (PM) region for magnetic fields parallel to all three principle axis is shown in Table ~\ref{tab:curie_weiss}. The effective magnetic moment determined experimentally is comparable to the theoretical free-ion value of 9.72 $\mu$$_{B}$ for the Tb$^{3+}$ ion. The negative $\theta$$_P$ indicates that antiferromagnetic exchange interactions dominate in the system. The difference in the calculated values of $\theta$$_P$ along different magnetic field directions, particularly along field parallel to a-axis, infers the role of strong magnetic anisotropy/single-ion anisotropy of Tb \cite{PhysRevB.100.094442, PhysRevB.99.100406}, due to different magnetic-site interactions present in the system. Next, to gain a deeper insight into the behavior of these anisotropic magnetic phase transitions, we have measured $\chi_{dc}$(\textit{T}) with different applied magnetic fields up to 14~T in all three directions. The $\chi_{dc}$(\textit{T}) plots with applied magnetic fields for $B \parallel a$ look distinctive, as shown in Fig.~\ref{fgr:MvsT}(a). With increasing magnetic field, $T_{N2}$ is rapidly suppressed and disappears above ${\sim}$5 T. In contrast, the higher-temperature transition $T_{N1}$ gradually shifts to lower temperatures with increasing field, reaching ${\sim}$24 K at 14 T. In addition, a new field-induced feature emerges, as observed in the 12 T susceptibility shown in Fig.~\ref{fgr:MvsT}(a). This transition, labeled $T^*$, is distinct from $T_{N2}$ and is only observed for $B \parallel a$. The field dependence of this phase transition is discussed in Fig.~\ref{fgr:Phase Diagram}.
There is not much variation in $T_{N1}$ and $T_{N2}$ for the cases $B \parallel b$ and $B \parallel c$ [Fig.~\ref{fgr:MvsT}(b) and (c)]. Additionally, the distorted triangular lattice of Tb may promote anisotropic magnetic interactions, contributing to the emergence of these intriguing magnetic phases. Note that the Néel temperature of TbAlGe ($T_{N1}$ = 40~K) is markedly higher than the value of $\sim$20~K expected from de Gennes scaling based on the Gd analog ($T_{N}$ = 30~K) \cite{r7gz-kwqw}. This enhancement indicates a breakdown of simple RKKY scaling, likely caused by strong anisotropic exchange interactions and crystal-field effects \cite{MAJUMDAR2001645}. Such behavior underscores the complex interplay between localized 4$f$ magnetism and itinerant electron topology in such a magnetic system.


\begin{table}[htbp]
\centering
{\caption{Curie--Weiss fitting parameters obtained from inverse magnetic susceptibility in the paramagnetic region for different field orientations.}
\setlength{\tabcolsep}{17pt}
\begin{tabular}{ccc}\\
\hline\hline
Field Direction & $\mu_{\mathrm{eff}}$ ($\mu_B$/Tb) & $\theta_P$ (K) \\
\hline
$B \parallel a$ & 10.19 & $-9.5$ \\
$B \parallel b$ & 9.78  & $-88.1$ \\
$B \parallel c$ & 9.86  & $-53.7$ \\
\hline\hline
\end{tabular}
\label{tab:curie_weiss}}
\end{table}
To further investigate the effect of the magnetic field, we have measured isothermal magnetization at selected temperatures up to 50 K. Fig.~\ref{fgr:MvsB}(a) shows magnetization as a function of the field at different temperatures between 2-35~K for B $\parallel$ a. An interesting observation is that, at the lowest measured temperature (2~K), significant hysteresis is observed on the slopes of the $M(B)$ curves near 3~T and 11~T. Such a first-order-like magnetic transition gradually gets smeared with increasing temperature. The transition observed around 11~T shifts to higher fields with increasing temperature, suggesting the presence of more intriguing magnetic behavior at further higher field values. The $M(B)$ curves for other directions are observed to be predominantly reversible with magnetic field directions without any field-induced transition. Moreover, $M(B)$ curves do not saturate even in a 14~T field at 2~K, and the spontaneous magnetic moment at 2~K is only ~4$\mu$$_{B}$/Tb$^{3+}$ (even lower for other directions), which is less than half of the free Tb$^{3+}$ magnetic moment. A comprehensive determination of the magnetic structure is necessary to clarify this behavior. Detailed inelastic neutron scattering experiments are necessary to directly elucidate the role of anisotropic exchange interactions in the observed anisotropic metamagnetic transitions. The suggested change in the magnetic structure at the lower transition temperature is further supported by the first-order magnetic phase transition observed near 8 K in TbAlGe, which may be associated with a change in the magnetic propagation vector driven by anisotropic exchange interactions. Moreover, the crystal structure reveals an isosceles triangular arrangement of Tb atoms along the a axis, perpendicular to the (0kl) planes. This geometrical motif likely underpins the unusual metamagnetic behavior observed exclusively for magnetic fields applied along the a-axis ($B\parallel a$).


\begin{figure}[htbp]
	\adjincludegraphics[height=14cm,trim={1cm 1cm 0cm 2cm},clip]{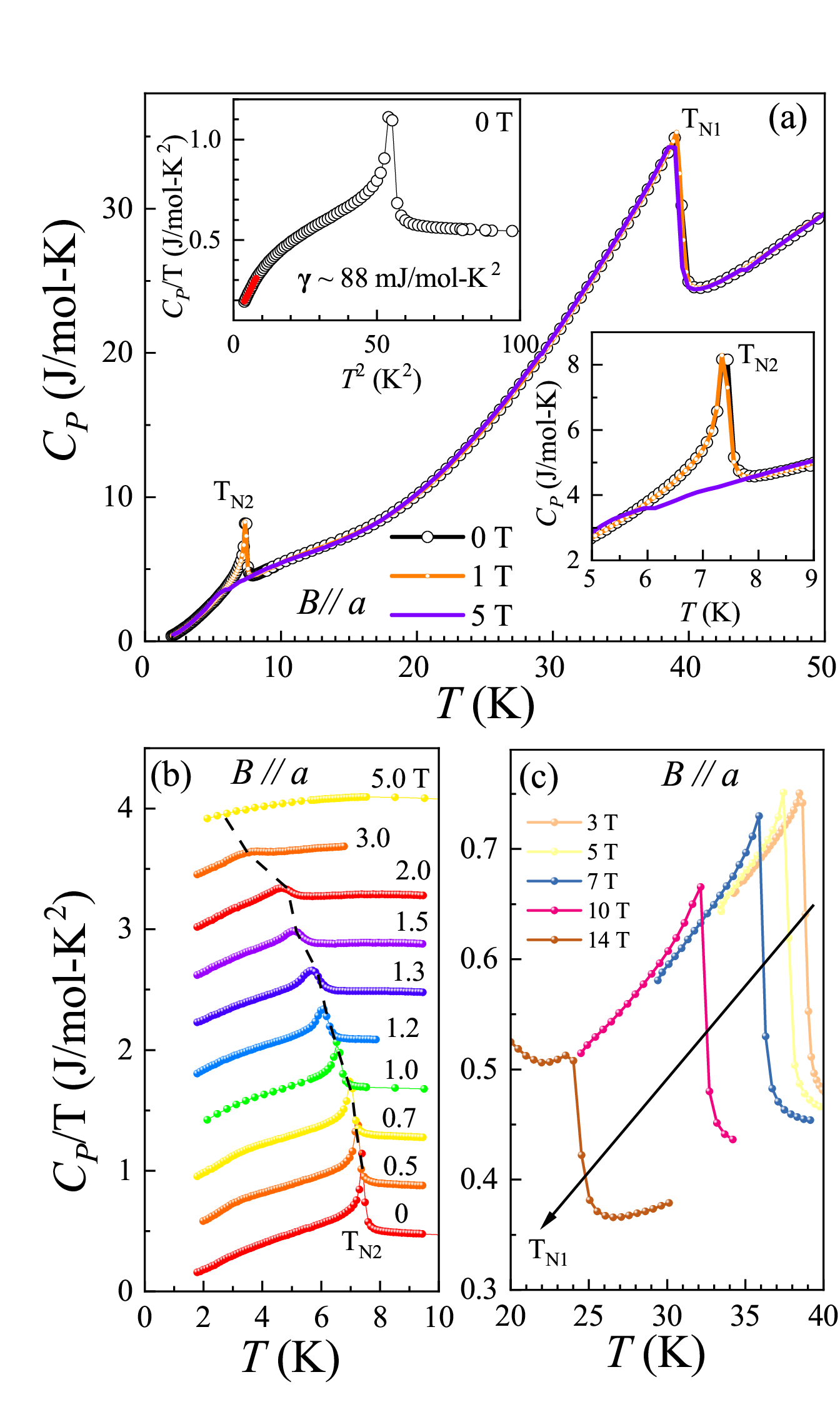}
    \caption{(a) Temperature dependence of Heat Capacity of TbAlGe measured under zero field and applied magnetic fields. The upper left inset shows the C$_P$/T vs T$^2$ plot in zero field with red solid line at low temperature region showing the liner fit to calculate the $\gamma$. The bottom right inset shows the suppression of $T_{N2}$ with an increase in the applied magnetic field. The C$_P$/T vs T plots with different applied magnetic fields for a better view of $T_{N1}$ and $T_{N2}$ are shown in fig.~(b) and (c), respectively.}
    \label{fgr:Cp}
 \end{figure}


 \begin{figure}[htbp]
\adjincludegraphics[height=9.5cm,trim={1cm 0cm 0.25cm 0.1cm},clip]{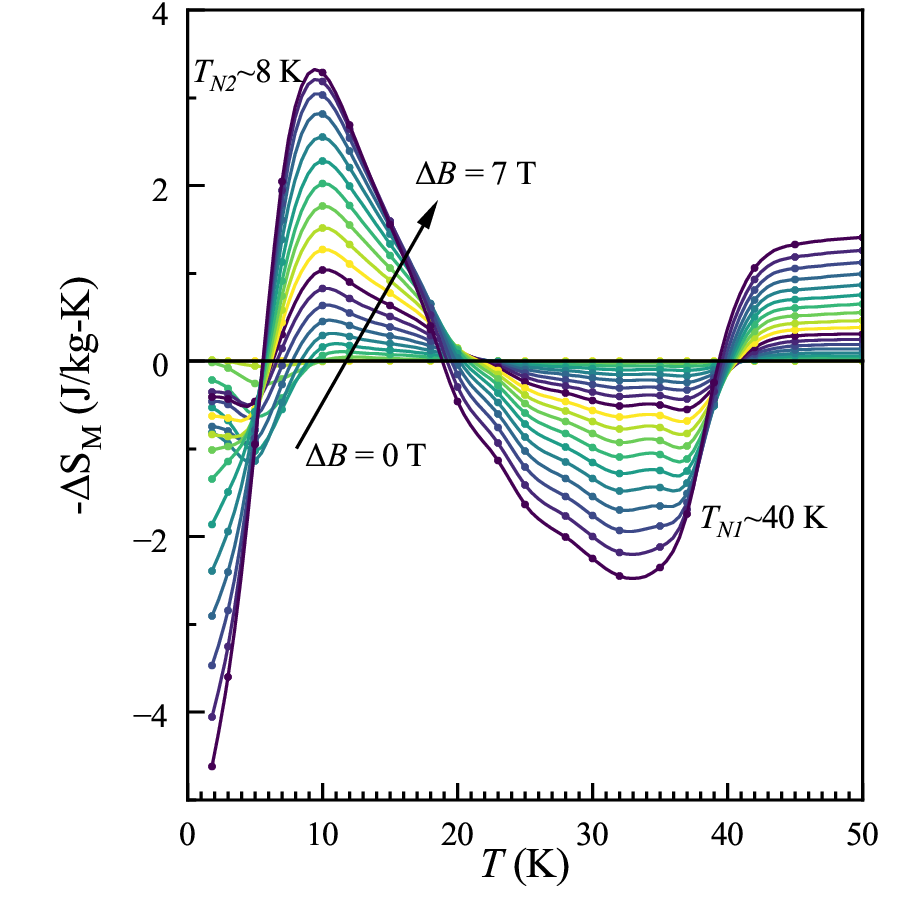} \caption{Change in magnetic entropy (-$\Delta$S) of TbAlGe as a function of temperature at different magnetic fields from 0 T to 7~T in 0.025~T steps. The observed sign change in entropy reveals possible competing magnetic phases that evolve with temperature and magnetic field.}
    \label{fgr:entropy}
 \end{figure}



\begin{figure*}[htbp]
	\adjincludegraphics[height=10cm,trim={2.3cm 1cm 0.8cm 0.5cm},clip]{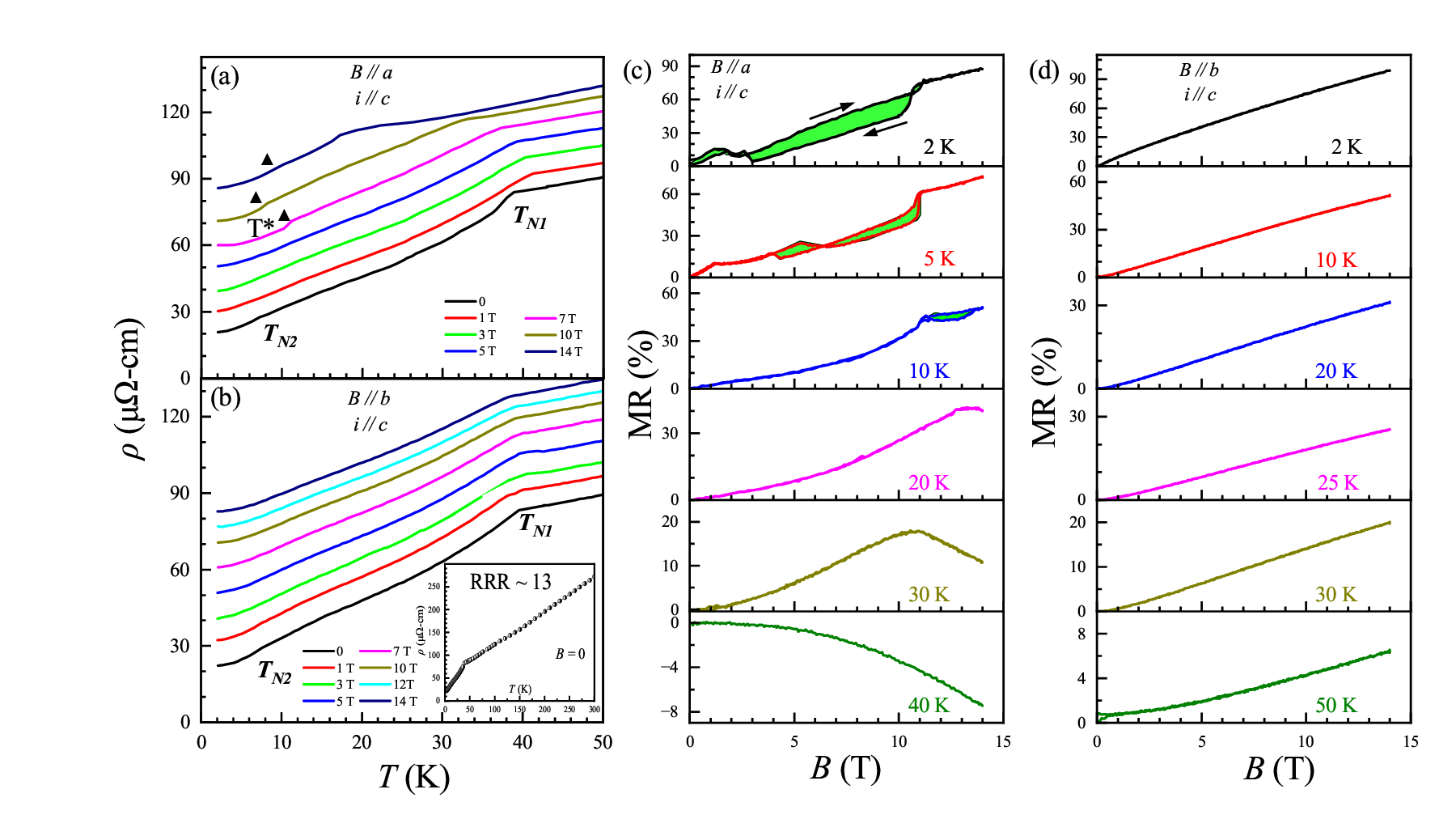}
    \caption{Electrical resistivity (${\rho}$) as a function of the temperature under different applied fields. (a) ${\rho}$$_{xx}$(T) of TbAlGe for $B \parallel a$, and $B \parallel b$ is shown in panels (a) and (b), respectively. The curves above the zero field one are constantly shifted upwards with a factor of 10 $\mu$$\Omega$cm  for the clarity of the magnetic transition under applied fields. Black triangles in (a) show the field-induced transition (\textit{T$^*$}). Inset of (b) shows the ${\rho}$$_{xx}$(T) up to 300 K. Transverse magnetoresistance of TbAlGe and its evolution with temperature for $B \parallel a$, and $B \parallel b$, are shown in Figs. (c), and (d), respectively.}
    
    \label{fgr:RT_RH}
 \end{figure*}


 \begin{figure}[htbp]
	\adjincludegraphics[height=9.5cm,trim={0.1cm 0cm 0.2cm 0.2cm},clip]{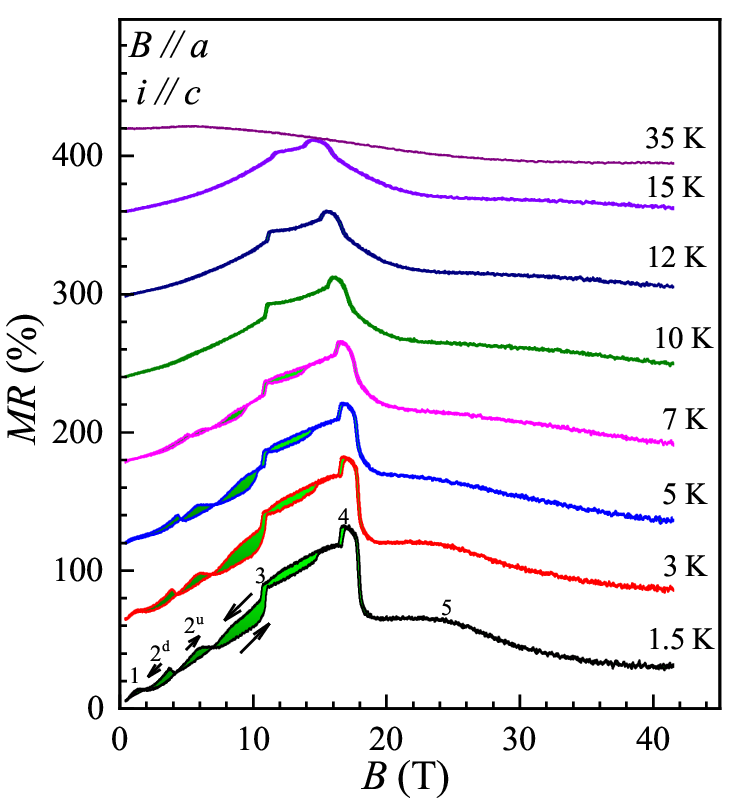}
    \caption{ High-field magnetoresistance as a function of the magnetic field at different temperatures in the magnetic state for $B \parallel a$. The curves for temperatures above 1.5~K are constantly shifted upward (with a factor of 60\%) for clarity to illustrate the multiple field-induced transitions as marked with different numbers ( with notation `u': up sweep, and `d': down sweep) symbols for 1.5~K curve.}
    \label{fgr:HF_MR}
 \end{figure}


   \begin{figure}[!htbp]
	\adjincludegraphics[height=12cm,trim={0.2cm 0.2cm 0cm 0cm},clip]{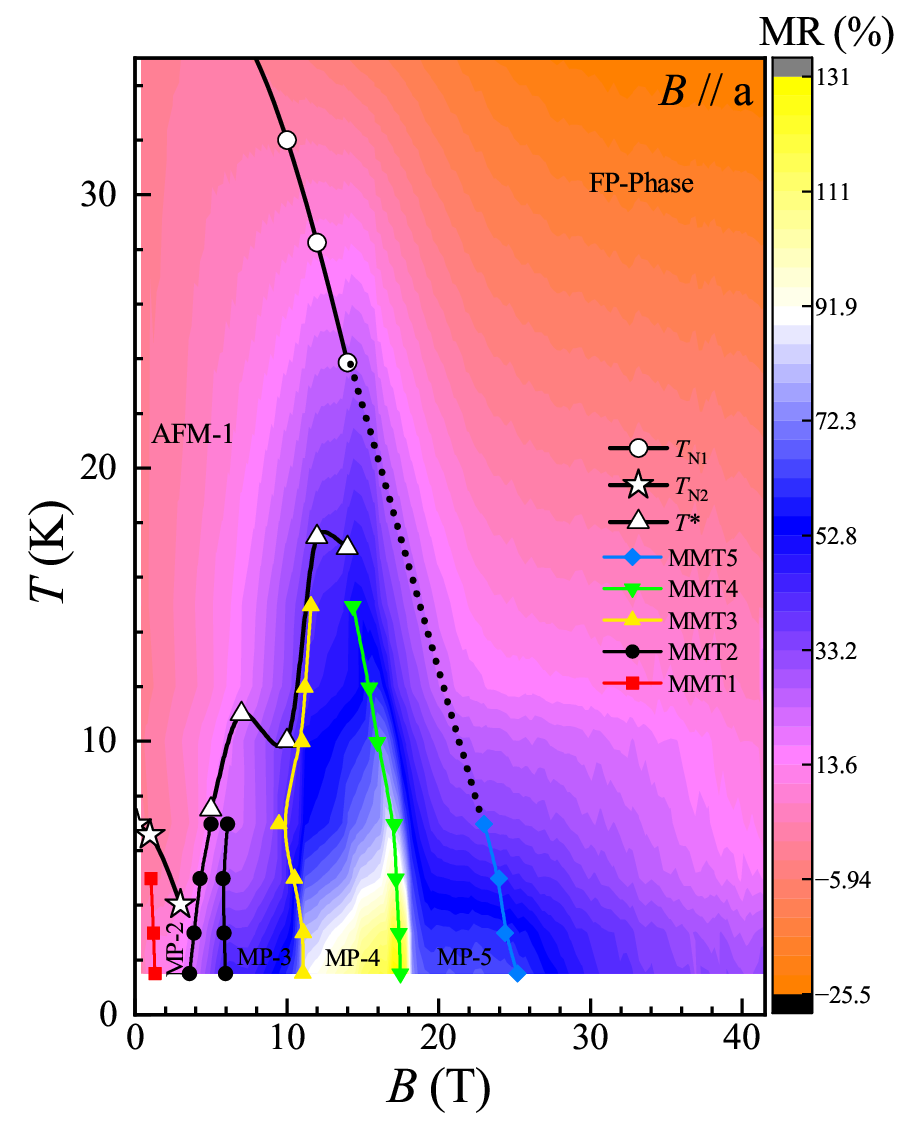}
    \caption{\textit{B}-\textit{T} phase diagram of TbAlGe as a contour plot of magnetoresistance up to 41.5~T, with  \textit{$T_{N1}$}, \textit{$T_{N2}$} and \textit{$T^{*}$} obtained from magnetic susceptibility measurements. The different metamagnetic transitions are derived from the magnetoresistance plot as shown in Fig.~\ref{fgr:HF_MR}.}
    \label{fgr:Phase Diagram}
 \end{figure}


\subsection{Heat Capacity and Magnetocaloric Effect}
We report the results of zero-field and in-field heat capacity, C$_{P}$(T), measurements of TbAlGe as a function of temperature, which is shown in Fig.~\ref{fgr:Cp}(a). There is a well-defined $\lambda$ anomaly at the \textit{$T_{N1}$}${\sim}$ 40~K, and another magnetic transition at a much lower temperature, \textit{$T_{N2}$}${\sim}$ 8~K, which is also seen in the magnetic measurements. Both the peaks at $\textit{$T_{N1}$}$ and \textit{$T_{N2}$} are gradually shifted towards lower temperatures with an increase in applied magnetic fields below 5 T. For \textit{B} =5~T (blue solid line), the \textit{$T_{N2}$} is smeared out, which shows how delicate the transition is compared to \textit{$T_{N1}$}, despite both being antiferromagnetic-type transitions. Such an interesting behavior might be due to a modulated AFM structure, which needs to be further investigated by neutron scattering. As calculated from the linear fitting of the low-temperature C$_P$/T(T$^2$) data of TbAlGe [as shown in the insets of Fig.~\ref{fgr:Cp}(a)], the Sommerfeld coefficient, $\gamma$= 88~mJ/mol~K$^2$, suggests strong electronic correlations. The lower panels of Fig.~\ref{fgr:Cp} show the C$_P$/T as a function of temperature, providing a clearer view of the effects of the applied magnetic field on these magnetic phase transitions in TbAlGe.

We have studied the magnetocaloric effect (MCE) or magnetic entropy change, -$\Delta$S$_M$ (the change induced in the entropy, S, by the application of \textit{B} at a given \textit{T}), in TbAlGe, particularly along the $a$-axis, which seems to be a peculiar one compared to other principal axes, to get more information about the different magnetic phase transitions, as studied in other Tb compounds \cite{PhysRevMaterials.5.054407, 10.1063/1.5121293, PhysRevB.101.144440, 10.1063/1.4903770, PhysRevB.80.134422, Fang2017, PhysRevB.65.094405}. We have measured many isothermal M(B) curves for different temperatures with close intervals of 2~K up to 50~K, to calculate the MCE. These values have been derived for an initial \textit{B} of zero and final \textit{B} marked in the figures. The $\Delta$S from M data is determined through the Maxwell relation, (\textit{$\partial$S}(\textit{T,B})/\textit{$\partial$H})$_T$=(\textit{$\partial$M}(\textit{T,B})/\textit{$\partial$T})$_B$  \cite {tishin2016magnetocaloric}. We have derived the isothermal entropy change, defined as -$\Delta$S$_M$, up to 50~K, and the result obtained is shown in Fig.~\ref{fgr:entropy}. The magnetic susceptibility data corroborate well with the two magnetic phase transitions revealed from magnetic entropy as well. In addition, the sign of -$\Delta$S$_M$ is flipped twice below the magnetic ordering, (i) around 20~K, exactly where $\chi$ shows a minima between the two transitions [Fig.~\ref{fgr:MvsT}(a)], indicates the strong competition among the magnetic interactions and implies the dominance of an antiferromagnetic component over the ferromagnetic (FM) phase \cite {PECHARSKY199944}, and (ii) at low temperature just below \textit{$T_{N2}$}, where we expect a possible first-order magnetic phase transition (also discussed in magnetization); at higher temperatures, the sign of -$\Delta$S$_M$ is positive, which is a signature of a tendency for spin-reorientation \cite {LI2020153810}, suggesting the dominance of ferromagnetic correlations in the presence of such magnitudes of external fields. The presence of a possible first-order magnetic transition below 8~K, manifested through magnetic hysteresis [Fig.~\ref{fgr:MvsB}] and corroborated by magnetotransport data [Figs.~\ref{fgr:RT_RH} and ~\ref{fgr:HF_MR}], together with multiple sign reversals of the magnetic entropy, points to significant magnetocaloric potential in this compound.

\subsection {Magnetotransport Properties up to high magnetic field of 41.5~T}
 We have measured detailed zero-field and in-field transverse electrical resistivity ${\rho}$ as a function of temperature in both directions, $B \parallel a$ and $B \parallel b$ directions, as shown in Fig.~\ref{fgr:RT_RH}(a) and (b), respectively. The residual resistivity ratio (RRR), calculated as ${\rho}$(300 K)/${\rho}$(1.8~K) and equal to about 13 (as shown in the inset of (b)), indicates fair crystal quality. The electrical resistivity of TbAlGe shows metallic behavior, but exhibits an anomaly as the system approaches \textit{$T_{N1}$} from higher temperatures. Below \textit{$T_{N1}$}, the loss of spin-disorder scattering due to long-range AFM ordering leads to a rapid decrease in resistivity. There is a clear, sharp drop in ${\rho}$ at \textit{$T_{N1}$} in the zero field resistivity data Fig.~\ref{fgr:RT_RH}(a) and (b). With further lowering of the temperature, there is a shoulder/hump-like feature near 8~K as the system enters another magnetic phase. These anomalies at the onset of magnetic ordering manifest differently depending on the magnetic field direction along different crystallographic directions. Similar to the magnetic susceptibility shown in Fig.~\ref{fgr:MvsT}(a), $T^*$ is also observed in the resistivity only for $B \parallel a$. To further explore the existing magnetic phases at higher fields, we have measured the magnetoresistance (MR) at several temperatures for \textit{$B \parallel a$} and $B \parallel b$ as shown in Fig.~\ref{fgr:RT_RH}(c) and (d), respectively. Having multiple metamagnetic transitions along the $a$-axis, similar to magnetization, confirms that $B \parallel a$ is the most favorable direction for observing emergent magnetotransport properties.
 
 The novel features observed in the magnetic and magnetotransport measurements motivate us to study further higher magnetic fields (\textit{B} = 41.5~T), mainly for the applied magnetic field along the $a$-axis. The MR behavior at different temperatures starting from base temperature, T = 1.5~K to 35 K (in the magnetic state) is shown in Fig.~\ref{fgr:HF_MR}. The symmetrized magnetoresistance data to remove the contribution from the Hall component using, MR(B) = [MR(B) + MR(-B)]/2 are shown here. The curves for temperatures above 1.5~K are constantly shifted upward (with a factor of 60\%) for clarity to illustrate the multiple field-induced transitions. Interestingly, a clear hysteresis is observed between the up-sweep and down-sweep magnetoresistance curves, but only below 10~K. This hysteretic behavior, consistent with that seen in magnetization, further reinforces the presence of a first-order magnetic phase transition in TbAlGe. Measurements in magnetic fields upto 41.5~T show an additional set of metamagnetic transitions beyond those detected in the PPMS measurements up to 14~T. These metamagnetic transitions are labeled sequentially from 1 to 5; for instance, the first metamagnetic transition is denoted as MMT1 (as plotted along with contour plot in Fig.8). In particular, the magnetoresistance reaches approximately 120\% at 1.5~K around 18~T. Notably, TbAlGe hosts both first-order and second-order magnetic phase transitions, consistent with the rich field-induced phase diagram (will discuss in the next section).

\subsection{Phase Diagram}

Our results can be summarized in the magnetic phase diagram of TbAlGe, measured down to 1.5~K and up to an applied magnetic field of 41.5~T. Contour plots of magnetoresistance (\textit{B} = 0–41.5~T), are shown in Fig.~\ref{fgr:Phase Diagram}, where the white data points mark the transition temperatures \textit{$T_{N1}$}, \textit{$T_{N2}$}, and \textit{$T^{*}$}, determined from the derivative of the magnetic susceptibility. The high-field magnetoresistance data presented in Fig.~\ref{fgr:HF_MR} are used to identify a series of metamagnetic transitions (MMTs) and to construct the phase diagram. At 1.5 K, five distinct metamagnetic transitions are observed in the range of 2–26 T, labeled as MMT1 through MMT5, separating multiple magnetic phases (MP-1 to MP-5). Among these, MMT2 exhibits two distinct shoulders (2$^u$ and 2$^d$) at different magnetic fields of up- and down-sweep in MR traces, indicating a first-order transition. In contrast, the remaining transitions show no such crossing; however, significant hysteresis is observed from the touching points of the up- and down-sweep MR curves. The shoulders in the up-sweep MR curves are used to further determine the phase boundaries of all the MMTs. With increasing temperature, the hysteresis associated with these MMTs becomes progressively narrower, as observed in Fig.~\ref{fgr:HF_MR}, suggesting a weakening of the first-order character. The resulting phase diagram consists of multiple magnetic phases separated by these transitions. The first antiferromagnetic (AFM) phase eventually merges with the fifth magnetic phase (MP-5) around 26~T, clearly delineates the boundary between the field-polarized (FP) phase and the lower-field phases.The phase diagram highlights the fragility of the magnetic phases between the two antiferromagnetic transitions (\textit{$T_{N1}$} and \textit{$T_{N2}$}), along with an additional field-induced transition (\textit{$T^{*}$}), revealing a complex interplay of competing magnetic interactions.There is strong agreement between transport and magnetic measurements, particularly in the overlapping temperature range. These phase transitions are also reflected in the magnetic entropy, where sign changes correspond to different magnetic phases. The sliding triangular arrangement of Tb atoms with distorted bond lengths may introduce moderate geometrical frustration.  The presence of anisotropic exchange interaction due to frustration gives rise to the possibility of exotic magnetism, as seen in other geometrically frustrated heavy rare-earth magnets as well \cite{PhysRevB.78.064412, PhysRevB.100.094442, PhysRevB.101.144440, 10.1063/1.5121293, 10.1103/physrevb.60.12162, kurumaji2019skyrmion, 1sc3-6cfp, annurev:/content/journals/10.1146/annurev.ms.24.080194.002321}. 
The resulting anisotropic exchange interactions likely play a key role in stabilizing the non-trivial spin structures in TbAlGe, which may host topological electronic states. 

\section{Conclusion}

In conclusion, we have grown single crystals of the heavy rare-earth system TbAlGe and performed detailed magnetic and magnetotransport studies on this metallic antiferromagnet, which crystallizes in the orthorhombic $Cmcm$ structure. TbAlGe exhibits two antiferromagnetic transitions at $T_{N1}$ = 40~K and $T_{N2}$ = 8~K, developing a rich, field-tuned magnetic phase diagram with multiple metamagnetic transitions at 1.5~K. Notably, unlike other heavy rare-earth RAlGe compounds, TbAlGe displays a strongly first-order transition near 6~T in the vicinity of the lower Néel temperature. High-field measurements up to 41.5~T reveal additional metamagnetic transitions beyond those observed in conventional measurements up to 14~T, highlighting sophisticated anisotropic magnetic interactions, likely influenced by the distorted triangular lattice of Terbium. These results provide key insight into the interplay between magnetism and Fermi-surface topology in TbAlGe, making it an ideal platform to study correlated electronic physics. Future studies combining band-structure calculations and neutron scattering experiments will be essential to fully elucidate the connection between the magnetic phases and the topological electronic structure in this system.

\begin{acknowledgments}
Research at the University of Maryland was supported by the Gordon and Betty Moore Foundation’s EPiQS Initiative through Grant No. GBMF9071, the U.S. National Science Foundation (NSF) Grant No. DMR2303090, the Binational Science Foundation Grant No. 2022126, and the Maryland Quantum Materials Center. R. K. acknowledges support from the NSF Grant Number 2201516 under the Accelnet program of Office of International Science and Engineering (OISE). A portion of this work was performed at the National High Magnetic Field Laboratory, which is supported by National Science Foundation Cooperative Agreement No. DMR-2128556 and the State of Florida. S.R.S. acknowledges support from the National Institute of Standards and Technology Cooperative Agreement 70NANB17H301.
\end{acknowledgments}

\nocite{*}

\bibliography{TbAlGe}

@Article{Bernevig2022,
author={Bernevig, B. Andrei
and Felser, Claudia
and Beidenkopf, Haim},
title={Progress and prospects in magnetic topological materials},
journal={Nature},
year={2022},
month={Mar},
day={01},
volume={603},
number={7899},
pages={41-51},
issn={1476-4687},
doi={10.1038/s41586-021-04105-x},
url={https://doi.org/10.1038/s41586-021-04105-x}
}

@article{KHJBuschow_1977,
doi = {10.1088/0034-4885/40/10/002},
url = {https://dx.doi.org/10.1088/0034-4885/40/10/002},
year = {1977},
month = {oct},
publisher = {},
volume = {40},
number = {10},
pages = {1179},
author = {K H J Buschow},
title = {Intermetallic compounds of rare-earth and 3d transition metals},
journal = {Reports on Progress in Physics},
}

@article{RevModPhys.81.1551,
  title = {Superconducting phases of $f$-electron compounds},
  author = {Pfleiderer, Christian},
  journal = {Rev. Mod. Phys.},
  volume = {81},
  issue = {4},
  pages = {1551--1624},
  numpages = {0},
  year = {2009},
  month = {Nov},
  publisher = {American Physical Society},
  doi = {10.1103/RevModPhys.81.1551},
  url = {https://link.aps.org/doi/10.1103/RevModPhys.81.1551}
}

@article{PhysRevB.84.235126,
  title = {Topological nodal semimetals},
  author = {Burkov, A. A. and Hook, M. D. and Balents, Leon},
  journal = {Phys. Rev. B},
  volume = {84},
  issue = {23},
  pages = {235126},
  numpages = {14},
  year = {2011},
  month = {Dec},
  publisher = {American Physical Society},
  doi = {10.1103/PhysRevB.84.235126},
  url = {https://link.aps.org/doi/10.1103/PhysRevB.84.235126}
}

@article{PhysRevB.105.174502,
  title = {Pressure-induced superconductivity in the noncentrosymmetric Weyl semimetals $\mathrm{LaAl}X$ $(X=\mathrm{Si},\mathrm{Ge})$},
  author = {Cao, Weizheng and Zhao, Ningning and Pei, Cuiying and Wang, Qi and Zhang, Qinghua and Ying, Tianping and Zhao, Yi and Gao, Lingling and Li, Changhua and Yu, Na and Gu, Lin and Chen, Yulin and Liu, Kai and Qi, Yanpeng},
  journal = {Phys. Rev. B},
  volume = {105},
  issue = {17},
  pages = {174502},
  numpages = {8},
  year = {2022},
  month = {May},
  publisher = {American Physical Society},
  doi = {10.1103/PhysRevB.105.174502},
  url = {https://link.aps.org/doi/10.1103/PhysRevB.105.174502}
}

@article{PhysRevB.103.165128,
  title = {Multiple Weyl fermions in the noncentrosymmetric semimetal LaAlSi},
  author = {Su, Hao and Shi, Xianbiao and Yuan, Jian and Wan, Yimin and Cheng, Erjian and Xi, Chuanying and Pi, Li and Wang, Xia and Zou, Zhiqiang and Yu, Na and Zhao, Weiwei and Li, Shiyan and Guo, Yanfeng},
  journal = {Phys. Rev. B},
  volume = {103},
  issue = {16},
  pages = {165128},
  numpages = {8},
  year = {2021},
  month = {Apr},
  publisher = {American Physical Society},
  doi = {10.1103/PhysRevB.103.165128},
  url = {https://link.aps.org/doi/10.1103/PhysRevB.103.165128}
}

@article{
doi:10.1126/sciadv.1603266,
author = {Su-Yang Xu  and Nasser Alidoust  and Guoqing Chang  and Hong Lu  and Bahadur Singh  and Ilya Belopolski  and Daniel S. Sanchez  and Xiao Zhang  and Guang Bian  and Hao Zheng  and Marious-Adrian Husanu  and Yi Bian  and Shin-Ming Huang  and Chuang-Han Hsu  and Tay-Rong Chang  and Horng-Tay Jeng  and Arun Bansil  and Titus Neupert  and Vladimir N. Strocov  and Hsin Lin  and Shuang Jia  and M. Zahid Hasan },
title = {Discovery of Lorentz-violating type II Weyl fermions in LaAlGe},
journal = {Science Advances},
volume = {3},
number = {6},
pages = {e1603266},
year = {2017},
doi = {10.1126/sciadv.1603266},
}

@article{PhysRevLett.124.017202,
  title = {Topological Magnetic Phase in the Candidate Weyl Semimetal CeAlGe},
  author = {Puphal, Pascal and Pomjakushin, Vladimir and Kanazawa, Naoya and Ukleev, Victor and Gawryluk, Dariusz J. and Ma, Junzhang and Naamneh, Muntaser and Plumb, Nicholas C. and Keller, Lukas and Cubitt, Robert and Pomjakushina, Ekaterina and White, Jonathan S.},
  journal = {Phys. Rev. Lett.},
  volume = {124},
  issue = {1},
  pages = {017202},
  numpages = {7},
  year = {2020},
  month = {Jan},
  publisher = {American Physical Society},
  doi = {10.1103/PhysRevLett.124.017202},
  url = {https://link.aps.org/doi/10.1103/PhysRevLett.124.017202}
}

@article{PhysRevB.103.214401,
  title = {Critical behavior of the magnetic Weyl semimetal PrAlGe},
  author = {Liu, Wei and Zhao, Jun and Meng, Fanying and Rahman, Azizur and Qin, Yongliang and Fan, Jiyu and Pi, Li and Tian, Zhaoming and Du, Haifeng and Zhang, Lei and Zhang, Yuheng},
  journal = {Phys. Rev. B},
  volume = {103},
  issue = {21},
  pages = {214401},
  numpages = {8},
  year = {2021},
  month = {Jun},
  publisher = {American Physical Society},
  doi = {10.1103/PhysRevB.103.214401},
  url = {https://link.aps.org/doi/10.1103/PhysRevB.103.214401}
}

@article{PhysRevB.97.041104,
  title = {Magnetic and noncentrosymmetric Weyl fermion semimetals in the $\mathit{R}\mathrm{AlGe}$ family of compounds ($\mathit{R}=\mathrm{rare}\phantom{\rule{0.28em}{0ex}}\mathrm{earth}$)},
  author = {Chang, Guoqing and Singh, Bahadur and Xu, Su-Yang and Bian, Guang and Huang, Shin-Ming and Hsu, Chuang-Han and Belopolski, Ilya and Alidoust, Nasser and Sanchez, Daniel S. and Zheng, Hao and Lu, Hong and Zhang, Xiao and Bian, Yi and Chang, Tay-Rong and Jeng, Horng-Tay and Bansil, Arun and Hsu, Han and Jia, Shuang and Neupert, Titus and Lin, Hsin and Hasan, M. Zahid},
  journal = {Phys. Rev. B},
  volume = {97},
  issue = {4},
  pages = {041104},
  numpages = {7},
  year = {2018},
  month = {Jan},
  publisher = {American Physical Society},
  doi = {10.1103/PhysRevB.97.041104},
  url = {https://link.aps.org/doi/10.1103/PhysRevB.97.041104}
}

@article{doi:10.1126/science.1245085,
author = {Z. K. Liu  and B. Zhou  and Y. Zhang  and Z. J. Wang  and H. M. Weng  and D. Prabhakaran  and S.-K. Mo  and Z. X. Shen  and Z. Fang  and X. Dai  and Z. Hussain  and Y. L. Chen },
title = {Discovery of a Three-Dimensional Topological Dirac Semimetal, Na<sub>3</sub>Bi},
journal = {Science},
volume = {343},
number = {6173},
pages = {864-867},
year = {2014},
doi = {10.1126/science.1245085},
URL = {https://www.science.org/doi/abs/10.1126/science.1245085},
}

@Article{Neupane2014,
author={Neupane, Madhab
and Xu, Su-Yang
and Sankar, Raman
and Alidoust, Nasser
and Bian, Guang
and Liu, Chang
and Belopolski, Ilya
and Chang, Tay-Rong
and Jeng, Horng-Tay
and Lin, Hsin
and Bansil, Arun
and Chou, Fangcheng
and Hasan, M. Zahid},
title={Observation of a three-dimensional topological Dirac semimetal phase in high-mobility Cd3As2},
journal={Nature Communications},
year={2014},
month={May},
day={07},
volume={5},
number={1},
pages={3786},
issn={2041-1723},
doi={10.1038/ncomms4786},
url={https://doi.org/10.1038/ncomms4786}
}

@article{doi:10.1021/ic00025a021,
author = {Guloy, Arnold M. and Corbett, John D.},
title = {Syntheses and structures of lanthanum germanide, LaGe2-x, and lanthanum aluminum germanide, LaAlGe: interrelationships among the .alpha.-ThSi2, .alpha.-GdSi2, and LaPtSi structure types},
journal = {Inorganic Chemistry},
volume = {30},
number = {25},
pages = {4789-4794},
year = {1991},
doi = {10.1021/ic00025a021},

URL = {https://doi.org/10.1021/ic00025a021},
}

@article{doi:10.1021/jpcc.1c07073,
author = {Wang, Cong and Guo, Yongquan and Wang, Tai},
title = {Correlation between Slanted Magnetic Structure and Electromagnetic Responses in the RAlGe (R = Tb and Er) System},
journal = {The Journal of Physical Chemistry C},
volume = {125},
number = {39},
pages = {21764-21770},
year = {2021},
doi = {10.1021/acs.jpcc.1c07073},
}

@article{Wang_2020,
doi = {10.1088/1674-1056/abad25},
url = {https://doi.org/10.1088/1674-1056/abad25},
year = {2020},
month = {dec},
publisher = {Chinese Physical Society and IOP Publishing Ltd},
volume = {29},
number = {12},
pages = {127502},
author = {Wang, Cong and Guo, Yong-Quan and Wang, Tai and Yang, Shuo-Wang},
title = {Crystal structure and electromagnetic responses of tetragonal GdAlGe},
journal = {Chinese Physics B}
}

@article{Gouda_2022,
doi = {10.1088/1742-6596/2164/1/012072},
url = {https://dx.doi.org/10.1088/1742-6596/2164/1/012072},
year = {2022},
month = {mar},
publisher = {IOP Publishing},
volume = {2164},
number = {1},
pages = {012072},
author = {Kaiki Gouda and Takashi Nishioka},
title = {Angular-field magnetic phase diagram of b-plane at 4 K of YAlGe-type TbAlGe with zigzag-chain},
journal = {Journal of Physics: Conference Series},
}

@article{PUKAS2004162,
title = {Crystal structures of the RAlSi and RAlGe compounds},
journal = {Journal of Alloys and Compounds},
volume = {367},
number = {1},
pages = {162-166},
year = {2004},
issn = {0925-8388},
doi = {https://doi.org/10.1016/j.jallcom.2003.08.031},
author = {Svitlana Pukas and Yuri Lutsyshyn and Mykola Manyako and Evgen Gladyshevskii},
}

@article{Klepp:a21087,
author = "Klepp, K. and Parth{\'{e}}, E.",
title = "{{\it R}PtSi phases ({\it R} = La, Ce, Pr, Nd, Sm and Gd) with an ordered ThSi${\sb 2}$ derivative structure}",
journal = "Acta Crystallographica Section B",
year = "1982",
volume = "38",
number = "4",
pages = "1105--1108",
month = "Apr",
doi = {10.1107/S056774088200507X},
url = {https://doi.org/10.1107/S056774088200507X},
}

@article{WANG2022163623,
title = {Dynamic evolution from positive to negative magnetoresistance of RAlGe (R = Dy, Ho) with disordered orthorhombic structure},
journal = {Journal of Alloys and Compounds},
volume = {902},
pages = {163623},
year = {2022},
issn = {0925-8388},
doi = {https://doi.org/10.1016/j.jallcom.2022.163623},
author = {Cong Wang and Yongquan Guo and Tai Wang}
}

@article{WANG2021167739,
title = {Magnetic and transport properties of orthorhombic rare-earth aluminum germanide GdAlGe},
journal = {Journal of Magnetism and Magnetic Materials},
volume = {526},
pages = {167739},
year = {2021},
issn = {0304-8853},
doi = {https://doi.org/10.1016/j.jmmm.2021.167739},
author = {Cong Wang and Yongquan Guo and Tai Wang}
}

@article{PhysRevB.98.245132,
  title = {Single-crystal investigation of the proposed type-II Weyl semimetal CeAlGe},
  author = {Hodovanets, H. and Eckberg, C. J. and Zavalij, P. Y. and Kim, H. and Lin, W.-C. and Zic, M. and Campbell, D. J. and Higgins, J. S. and Paglione, J.},
  journal = {Phys. Rev. B},
  volume = {98},
  issue = {24},
  pages = {245132},
  numpages = {8},
  year = {2018},
  month = {Dec},
  publisher = {American Physical Society},
  doi = {10.1103/PhysRevB.98.245132},
  url = {https://link.aps.org/doi/10.1103/PhysRevB.98.245132}
}

@article{FLANDORFER1998191,
title = {The Systems Ce–Al–(Si, Ge): Phase Equilibria and Physical Properties},
journal = {Journal of Solid State Chemistry},
volume = {137},
number = {2},
pages = {191-205},
year = {1998},
issn = {0022-4596},
doi = {https://doi.org/10.1006/jssc.1997.7660},
url = {https://www.sciencedirect.com/science/article/pii/S002245969797660X},
author = {H. Flandorfer and D. Kaczorowski and J. Gröbner and P. Rogl and R. Wouters and C. Godart and A. Kostikas},
}

@article{
doi:10.1126/science.aat0348,
author = {T. Suzuki  and L. Savary  and J.-P. Liu  and J. W. Lynn  and L. Balents  and J. G. Checkelsky },
title = {Singular angular magnetoresistance in a magnetic nodal semimetal},
journal = {Science},
volume = {365},
number = {6451},
pages = {377-381},
year = {2019},
doi = {10.1126/science.aat0348},
URL = {https://www.science.org/doi/abs/10.1126/science.aat0348},
}

@article{Singh02072020,
author = {Karan Singh and K. Mukherjee},
title = {Spin–lattice relaxation phenomena in the magnetic state of a suggested Weyl semimetal CeAlGe},
journal = {Philosophical Magazine},
volume = {100},
number = {13},
pages = {1771--1787},
year = {2020},
publisher = {Taylor \& Francis},
doi = {10.1080/14786435.2020.1728588},
URL = { https://doi.org/10.1080/14786435.2020.1728588
},
}

@Article{Destraz2020,
author={Destraz, Daniel
and Das, Lakshmi
and Tsirkin, Stepan S.
and Xu, Yang
and Neupert, Titus
and Chang, J.
and Schilling, A.
and Grushin, Adolfo G.
and Kohlbrecher, Joachim
and Keller, Lukas
and Puphal, Pascal
and Pomjakushina, Ekaterina
and White, Jonathan S.},
title={Magnetism and anomalous transport in the Weyl semimetal PrAlGe: possible route to axial gauge fields},
journal={npj Quantum Materials},
year={2020},
month={Jan},
day={17},
volume={5},
number={1},
pages={5},
issn={2397-4648},
doi={10.1038/s41535-019-0207-7},
url={https://doi.org/10.1038/s41535-019-0207-7}
}

@Article{Sanchez2020,
author={Sanchez, Daniel S.
and Chang, Guoqing
and Belopolski, Ilya
and Lu, Hong
and Yin, Jia-Xin
and Alidoust, Nasser
and Xu, Xitong
and Cochran, Tyler A.
and Zhang, Xiao
and Bian, Yi
and Zhang, Songtian S.
and Liu, Yi-Yuan
and Ma, Jie
and Bian, Guang
and Lin, Hsin
and Xu, Su-Yang
and Jia, Shuang
and Hasan, M. Zahid},
title={Observation of Weyl fermions in a magnetic non-centrosymmetric crystal},
journal={Nature Communications},
year={2020},
month={Jul},
day={03},
volume={11},
number={1},
pages={3356},
issn={2041-1723},
doi={10.1038/s41467-020-16879-1},
url={https://doi.org/10.1038/s41467-020-16879-1}
}

@article{10.1063/1.5132958,
    author = {Yang, Hung-Yu and Singh, Bahadur and Lu, Baozhu and Huang, Cheng-Yi and Bahrami, Faranak and Chiu, Wei-Chi and Graf, David and Huang, Shin-Ming and Wang, Baokai and Lin, Hsin and Torchinsky, Darius and Bansil, Arun and Tafti, Fazel},
    title = {Transition from intrinsic to extrinsic anomalous Hall effect in the ferromagnetic Weyl semimetal PrAlGe1−xSix},
    journal = {APL Materials},
    volume = {8},
    number = {1},
    pages = {011111},
    year = {2020},
    month = {01},
    }

@article{Zhao_2022,
doi = {10.1088/1367-2630/ac430a},
url = {https://dx.doi.org/10.1088/1367-2630/ac430a},
year = {2022},
month = {jan},
publisher = {IOP Publishing},
volume = {24},
number = {1},
pages = {013010},
author = {Zhao, Jun and Liu, Wei and Rahman, Azizur and Meng, Fanying and Ling, Langsheng and Xi, Chuanying and Tong, Wei and Bai, Yuming and Tian, Zhaoming and Zhong, Yunbo and Hu, Ying and Pi, Li and Zhang, Lei and Zhang, Yuheng},
title = {Field-induced tricritical phenomenon and magnetic structures in magnetic Weyl semimetal candidate NdAlGe},
journal = {New Journal of Physics},
}

@article{PhysRevB.107.224414,
  title = {Multi-$k$ magnetic structure and large anomalous Hall effect in candidate magnetic Weyl semimetal NdAlGe},
  author = {Dhital, C. and Dally, R. L. and Ruvalcaba, R. and Gonzalez-Hernandez, R. and Guerrero-Sanchez, J. and Cao, H. B. and Zhang, Q. and Tian, W. and Wu, Y. and Frontzek, M. D. and Karna, S. K. and Meads, A. and Wilson, B. and Chapai, R. and Graf, D. and Bacsa, J. and Jin, R. and DiTusa, J. F.},
  journal = {Phys. Rev. B},
  volume = {107},
  issue = {22},
  pages = {224414},
  numpages = {16},
  year = {2023},
  month = {Jun},
  publisher = {American Physical Society},
  }

@article{PhysRevMaterials.7.034202,
  title = {Stripe helical magnetism and two regimes of anomalous Hall effect in NdAlGe},
  author = {Yang, Hung-Yu and Gaudet, Jonathan and Verma, Rahul and Baidya, Santu and Bahrami, Faranak and Yao, Xiaohan and Huang, Cheng-Yi and DeBeer-Schmitt, Lisa and Aczel, Adam A. and Xu, Guangyong and Lin, Hsin and Bansil, Arun and Singh, Bahadur and Tafti, Fazel},
  journal = {Phys. Rev. Mater.},
  volume = {7},
  issue = {3},
  pages = {034202},
  numpages = {14},
  year = {2023},
  month = {Mar},
  publisher = {American Physical Society},
  doi = {10.1103/PhysRevMaterials.7.034202},
  url = {https://link.aps.org/doi/10.1103/PhysRevMaterials.7.034202}
}

@article{10.1063/1.5121293,
    author = {Kumar, Ram and Iyer, Kartik K. and Paulose, P. L. and Sampathkumaran, E. V.},
    title = "{Spin-glass features at multiple temperatures and transport anomalies in Tb4PtAl}",
    journal = {Journal of Applied Physics},
    volume = {126},
    number = {12},
    pages = {123906},
    year = {2019},
    month = {09},
    issn = {0021-8979},
    doi = {10.1063/1.5121293},
    }

@article{PhysRevMaterials.5.054407,
  title = {Competing magnetic interactions and magnetoresistance anomalies in cubic intermetallic compounds, ${\mathrm{Gd}}_{4}\mathrm{RhAl}$ and ${\mathrm{Tb}}_{4}\mathrm{RhAl}$, and enhanced magnetocaloric effect for the Tb case},
  author = {Kumar, Ram and Iyer, Kartik K. and Paulose, P. L. and Sampathkumaran, E. V.},
  journal = {Phys. Rev. Mater.},
  volume = {5},
  issue = {5},
  pages = {054407},
  numpages = {10},
  year = {2021},
  month = {May},
  publisher = {American Physical Society},
  doi = {10.1103/PhysRevMaterials.5.054407},
  url = {https://link.aps.org/doi/10.1103/PhysRevMaterials.5.054407}
}

@article{PhysRevB.101.144440,
  title = {Magnetic and transport anomalies in ${R}_{2}\mathrm{RhS}{\mathrm{i}}_{3}\phantom{\rule{4pt}{0ex}}(R=\mathrm{Gd}$, Tb, and Dy) resembling those of the exotic magnetic material $\mathrm{G}{\mathrm{d}}_{2}\mathrm{PdS}{\mathrm{i}}_{3}$},
  author = {Kumar, Ram and Iyer, Kartik K. and Paulose, P. L. and Sampathkumaran, E. V.},
  journal = {Phys. Rev. B},
  volume = {101},
  issue = {14},
  pages = {144440},
  numpages = {7},
  year = {2020},
  month = {Apr},
  publisher = {American Physical Society},
  doi = {10.1103/PhysRevB.101.144440},
  url = {https://link.aps.org/doi/10.1103/PhysRevB.101.144440}
}

@book{tishin2016magnetocaloric,
  title={The magnetocaloric effect and its applications},
  author={Tishin, Aleksandr M and Spichkin, Youry I},
  year={2016},
  publisher={CRC Press}
}

@Article{Fang2017,
author={Fang, Chunsheng
and Li, Guoxing
and Wang, Jianli
and Hutchison, W. D.
and Ren, Q. Y.
and Deng, Zhenyan
and Ma, Guohong
and Dou, Shixue
and Campbell, S. J.
and Cheng, Zhenxiang},
title={New insight into magneto-structural phase transitions in layered TbMn2Ge2-based compounds},
journal={Scientific Reports},
year={2017},
month={Apr},
day={04},
volume={7},
number={1},
pages={45814},
issn={2045-2322},
doi={10.1038/srep45814},
url={https://doi.org/10.1038/srep45814}
}

@article{Rietveld:a07067,
author = "Rietveld, H. M.",
title = "{A profile refinement method for nuclear and magnetic structures}",
journal = "Journal of Applied Crystallography",
year = "1969",
volume = "2",
number = "2",
pages = "65--71",
month = "Jun",
doi = {10.1107/S0021889869006558},
url = {https://doi.org/10.1107/S0021889869006558},
}

@article{PECHARSKY199944,
title = {Magnetocaloric effect and magnetic refrigeration},
journal = {Journal of Magnetism and Magnetic Materials},
volume = {200},
number = {1},
pages = {44-56},
year = {1999},
issn = {0304-8853},
doi = {https://doi.org/10.1016/S0304-8853(99)00397-2},
url = {https://www.sciencedirect.com/science/article/pii/S0304885399003972},
author = {Vitalij K. Pecharsky and Karl A. {Gschneidner Jr}}
}

@article{LI2020153810,
title = {Recent progresses in exploring the rare earth based intermetallic compounds for cryogenic magnetic refrigeration},
journal = {Journal of Alloys and Compounds},
volume = {823},
pages = {153810},
year = {2020},
issn = {0925-8388},
doi = {https://doi.org/10.1016/j.jallcom.2020.153810},
url = {https://www.sciencedirect.com/science/article/pii/S0925838820301730},
author = {Lingwei Li and Mi Yan},
}

@article{PhysRevB.78.064412,
  title = {Field-induced antiferromagnetism and competition in the metamagnetic state of terbium gallium garnet},
  author = {Kamazawa, K. and Louca, Despina and Morinaga, R. and Sato, T. J. and Huang, Q. and Copley, J. R. D. and Qiu, Y.},
  journal = {Phys. Rev. B},
  volume = {78},
  issue = {6},
  pages = {064412},
  numpages = {5},
  year = {2008},
  month = {Aug},
  publisher = {American Physical Society},
  doi = {10.1103/PhysRevB.78.064412},
  url = {https://link.aps.org/doi/10.1103/PhysRevB.78.064412}
}

@article{PhysRevB.100.094442,
  title = {Magnetic order and single-ion anisotropy in ${\mathrm{Tb}}_{3}{\mathrm{Ga}}_{5}{\mathrm{O}}_{12}$},
  author = {Wawrzy\ifmmode \acute{n}\else \'{n}\fi{}czak, R. and Tomasello, B. and Manuel, P. and Khalyavin, D. and Le, M. D. and Guidi, T. and Cervellino, A. and Ziman, T. and Boehm, M. and Nilsen, G. J. and Fennell, T.},
  journal = {Phys. Rev. B},
  volume = {100},
  issue = {9},
  pages = {094442},
  numpages = {13},
  year = {2019},
  month = {Sep},
  publisher = {American Physical Society},
  doi = {10.1103/PhysRevB.100.094442},
  url = {https://link.aps.org/doi/10.1103/PhysRevB.100.094442}
}

@article{PhysRevB.99.100406,
  title = {Existence of a critical canting angle of magnetic moments to induce multiferroicity in the Haldane spin-chain system ${\mathrm{Tb}}_{2}{\mathrm{BaNiO}}_{5}$},
  author = {Kumar, Ram and Rayaprol, Sudhindra and Rajput, Sarita and Maitra, Tulika and Adroja, D. T. and Iyer, Kartik K. and Upadhyay, Sanjay K. and Sampathkumaran, E. V.},
  journal = {Phys. Rev. B},
  volume = {99},
  issue = {10},
  pages = {100406},
  numpages = {6},
  year = {2019},
  month = {Mar},
  publisher = {American Physical Society},
  doi = {10.1103/PhysRevB.99.100406},
  url = {https://link.aps.org/doi/10.1103/PhysRevB.99.100406}
}

@article{10.1103/physrevb.60.12162,
  author = {Saha, S. and Sugawara, H. and Matsuda, T. D. and Sato, H. and Mallik, R. C. and Sampathkumaran, E. V.},
  title = {Magnetic anisotropy, first-order-like metamagnetic transitions, and large negative magnetoresistance in single-crystalgd2pdsi3},
  journal = {Physical Review B},
  year = {1999},
  volume = {60},
  issue = {17},
  pages = {12162-12165},
  doi = {10.1103/physrevb.60.12162}
}

@article{kurumaji2019skyrmion,
  title={Skyrmion lattice with a giant topological Hall effect in a frustrated triangular-lattice magnet},
  author={Kurumaji, Takashi and Nakajima, Taro and Hirschberger, Max and Kikkawa, Akiko and Yamasaki, Yuichi and Sagayama, Hajime and Nakao, Hironori and Taguchi, Yasujiro and Arima, Taka-hisa and Tokura, Yoshinori},
  journal={Science},
  volume={365},
  number={6456},
  pages={914--918},
  year={2019},
  publisher={American Association for the Advancement of Science}
}

@article{1sc3-6cfp,
  title = {Complex single-site magnetism and magnetotransport in single-crystalline ${\mathrm{Gd}}_{2}{\mathrm{AlSi}}_{3}$},
  author = {Kumar, Ram and Saha, Shanta R. and Horn, Jarryd and Ikeda, A. and Sokratov, Danila and Anand, Yash and Saraf, Prathum and Dorman, Ryan and Hemley, E. and Iyer, K. K. and Paglione, Johnpierre},
  journal = {Phys. Rev. B},
  volume = {111},
  issue = {21},
  pages = {214426},
  numpages = {9},
  year = {2025},
  month = {Jun},
  publisher = {American Physical Society},
  doi = {10.1103/1sc3-6cfp},
  url = {https://link.aps.org/doi/10.1103/1sc3-6cfp}
}

@article{PhysRevB.65.094405,
  title = {Magnetic and structural phase diagram of ${\mathrm{Tb}}_{5}({\mathrm{Si}}_{x}{\mathrm{Ge}}_{1\ensuremath{-}x}{)}_{4}$},
  author = {Ritter, C. and Morellon, L. and Algarabel, P. A. and Magen, C. and Ibarra, M. R.},
  journal = {Phys. Rev. B},
  volume = {65},
  issue = {9},
  pages = {094405},
  numpages = {10},
  year = {2002},
  month = {Feb},
  publisher = {American Physical Society},
  doi = {10.1103/PhysRevB.65.094405},
  url = {https://link.aps.org/doi/10.1103/PhysRevB.65.094405}
}

@article{PhysRevB.80.134422,
  title = {Single-crystal neutron diffraction study of short-range magnetic correlations in ${\text{Tb}}_{5}{\text{Ge}}_{4}$},
  author = {Tian, W. and Kreyssig, A. and Zarestky, J. L. and Tan, L. and Nandi, S. and Goldman, A. I. and Lograsso, T. A. and Schlagel, D. L. and Gschneidner, K. A. and Pecharsky, V. K. and McQueeney, R. J.},
  journal = {Phys. Rev. B},
  volume = {80},
  issue = {13},
  pages = {134422},
  numpages = {5},
  year = {2009},
  month = {Oct},
  publisher = {American Physical Society},
  doi = {10.1103/PhysRevB.80.134422},
  url = {https://link.aps.org/doi/10.1103/PhysRevB.80.134422}
}

@article{10.1063/1.4903770,
    author = {Maji, Bibekananda and Ray, Mayukh K. and Suresh, K. G. and Banerjee, S.},
    title = {Large exchange bias and magnetocaloric effect in TbMn2Si2},
    journal = {Journal of Applied Physics},
    volume = {116},
    number = {21},
    pages = {213913},
    year = {2014},
    month = {12},
    issn = {0021-8979},
    doi = {10.1063/1.4903770},
    url = {https://doi.org/10.1063/1.4903770},
  }

@article{r7gz-kwqw,
  title = {Single-crystalline orthorhombic GdAlGe as a rare-earth magnetic Dirac nodal-line metal},
  author = {Laha, Antu and Yao, Juntao and Kundu, Asish K. and Aryal, Niraj and Rajapitamahuni, Anil and Vescovo, Elio and Camino, Fernando and Kisslinger, Kim and Zhang, Lihua and Nykypanchuk, Dmytro and Sears, J. and Tranquada, J. M. and Yin, Weiguo and Li, Qiang},
  journal = {Phys. Rev. B},
  volume = {112},
  issue = {12},
  pages = {125155},
  numpages = {12},
  year = {2025},
  month = {Sep},
  publisher = {American Physical Society},
  doi = {10.1103/r7gz-kwqw},
  url = {https://link.aps.org/doi/10.1103/r7gz-kwqw}
}

@article{annurev:/content/journals/10.1146/annurev.ms.24.080194.002321,
   author = "Ramirez, A P",
   title = "Strongly Geometrically Frustrated Magnets", 
   journal= "Annual Review of Materials Research",
   year = "1994",
   volume = "24",
   number = "Volume 24, ",
   pages = "453-480",
   doi = "https://doi.org/10.1146/annurev.ms.24.080194.002321",
   url = "https://www.annualreviews.org/content/journals/10.1146/annurev.ms.24.080194.002321",
   publisher = "Annual Reviews",
   issn = "1545-4118",
   type = "Journal Article",
  }

@article{MAJUMDAR2001645,
title = {Multiple magnetic transitions and anomalous magnetism in Tb2CuGe3},
journal = {Solid State Communications},
volume = {117},
number = {11},
pages = {645-648},
year = {2001},
issn = {0038-1098},
doi = {https://doi.org/10.1016/S0038-1098(01)00011-4},
url = {https://www.sciencedirect.com/science/article/pii/S0038109801000114},
author = {S. Majumdar and E.V. Sampathkumaran}
}

@article{https://doi.org/10.1002/pssa.2210450257,
author = {Gignoux, D. and Asmat, H.},
title = {Magnetic properties and structures of TbAlGa and HoAlGa},
journal = {physica status solidi (a)},
volume = {45},
number = {2},
pages = {K149-K153},
doi = {https://doi.org/10.1002/pssa.2210450257},
year = {1978}
}

\end{document}